\documentclass[journal=jpcbfk,manuscript=article]{achemso}
\SectionNumbersOn

\usepackage{chemformula} 
\usepackage[T1]{fontenc} 
\usepackage{graphicx}
\usepackage{dcolumn}
\usepackage{bm}
\usepackage[utf8]{inputenc}
\usepackage{etoolbox}
\usepackage[T1] {fontenc}
\usepackage{xcolor}
\usepackage{datetime}
\usepackage{geometry}
\usepackage{amsmath,amsfonts,amssymb}
\usepackage{fancyhdr}
\usepackage{subcaption}
\graphicspath{ {../} }
\usepackage{float}
\usepackage{mathtools}

\newcommand{\oo}[0]{\boldsymbol{\omega}}

\newcommand{\dd}[0]{\mathrm{d}}
\newcommand{\ii}[0]{\mathrm{i}}

\newcommand{\rr}[0]{\boldsymbol{r}}

\author{Mohamed Houssein Mohamed}
\affiliation[UL]{Université de Lorraine and CNRS, LPCT UMR 7019, F-54000, France}
\author{Odette Tannous}
\affiliation[UL]{Université de Lorraine and CNRS, LPCT UMR 7019, F-54000, France}
\author{Camille Muller}
\affiliation[UL]{Université de Lorraine and CNRS, LPCT UMR 7019, F-54000, France}
\author{Daniel Borgis}
\affiliation[ENS]{PASTEUR, Département de Chimie, École Normale Supérieure, PSL University, Sorbonne Université, CNRS, Paris 75005, France}
\author{Francesca Ingrosso}
\affiliation[UL]{Université de Lorraine and CNRS, LPCT UMR 7019, F-54000, France}
\author{Luc Belloni}
\affiliation[CEA]{LIONS, NIMBE, CEA, CNRS, Université Paris-Saclay, Gif-sur-Yvette 91191, France}
\author{Antoine Carof}
\affiliation[UL]{Université de Lorraine and CNRS, LPCT UMR 7019, F-54000, France}
\email{antoine.carof@univ-lorraine.fr}

\title{Accurate Solvation Properties in supercritical CO$_2$ with Molecular Density Functional Theory}

\begin{document}

\begin{abstract}
    Supercritical CO$_2$ is a highly efficient solvent for the development of more environmentally benign chemical processes. It is crucial to predict its solvation properties---the solvation free energy and the solvation structure---both accurately and at low computational cost. We show here that classical density functional theory (cDFT) can reproduce the solvation properties obtained from conventional molecular simulations, while requiring a computational effort that is several orders of magnitude lower. This excellent agreement is achieved using a molecular cDFT formalism based on a density that depends on both the positions and orientations of CO$_2$ molecules in the vicinity of the solute. We further examine several levels of approximation for the excess free-energy functional in cDFT and demonstrate that the homogeneous reference fluid approximation is sufficient to recover the molecular dynamics (MD) benchmark results. These findings open the way to extending molecular cDFT to other thermodynamic conditions.
\end{abstract}

\begin{figure}[!h]
    \centering
    \includegraphics[width=0.8\textwidth]{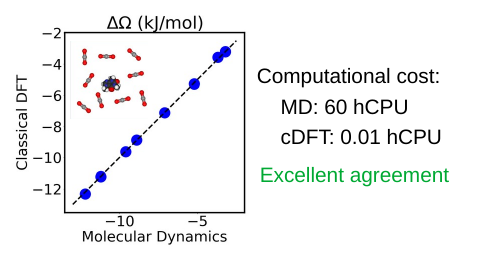}
    \caption*{TOC Graphic}
    \label{fig:toc}
\end{figure}

\section{Introduction}
\label{sec:introduction}

Supercritical CO$_2$ (scCO$_2$) is an environmentally benign solvent that exhibits physicochemical properties of considerable interest for a broad spectrum of industrial applications.~\cite{anastas_green_2010} It is non-toxic to humans and ecosystems and non-flammable, thereby constituting a ``greener'' alternative to conventional organic solvents.~\cite{montalban_supercritical_2022} The supercritical state of CO$_2$ is attained under relatively mild thermodynamic conditions, with a critical point at $P_c = 73.8$~bar, $T_c = 304$~K, and $n_c = 10.6$~mol.L$^{-1}$.~\cite{span_new_1996} In this regime, its physical and chemical properties gradually change from liquid-like and gas-like behaviour, and can be modulated in a controlled manner by small adjustments in pressure. In particular, fine control of the density of supercritical CO$_2$ enables precise tuning of its solvent power.~\cite{liu_physically_2010} Simultaneously, its gas-like character ensures enhanced mass transport and reduced viscosity compared to the liquid phase.  These combined features underpin a wide range of applications, particularly in the chemical industries. Supercritical CO$_2$ is widely employed as a selective solvent, capable of preferentially solvating specific solutes. Representative applications include the extraction of target compounds from natural matrices (\textit{e.g.}, the production of decaffeinated coffee),~\cite{reverchon_supercritical_1997, zosel_separation_1978} the impregnation of polymers for biomedical uses,~\cite{coutinho_aspirin-loaded_2021, weinstein_liquid_2010} supercritical fluid chromatography,~\cite{lesellier_many_2015, molineau_chromatographic_2021} and the separation of racemic mixtures.~\cite{johnston_separation_1987, tai_modified_2000} Furthermore, the unique properties of supercritical CO$_2$ are exploited in nanoparticle synthesis (\textit{e.g.}, drug-delivery systems) via micronisation techniques such as the supercritical anti-solvent process, expansion of supercritical solutions,~\cite{turk_manufacture_2009} and particles formation from gas-saturated solutions.~\cite{rossmann_solute_2012, sampaio_de_sousa_preparation_2007, turk_manufacture_2009}

The development and optimization of these applications necessitate a comprehensive understanding of the physicochemical properties of the scCO$_2$ fluid in interaction with solutes, under varying pressure and temperature conditions and in the presence of complex environments (\textit{e.g.}, confinement, interfaces, cosolvent). The chemical industry has established several parametric methods to predict solubility in scCO$_2$: (i) the equation-of-state (EoS) approach,~\cite{colussi_comparison_2006, garlapati_temperature_2009} in which the solubility is inferred from an EoS describing the mixtures (CO$_2$ and the solute and/or cosolvent); (ii) the density-based approach, which correlates solubility directly with the solvent density by means of semi-empirical expressions;~\cite{bartle_solubilities_1991, chrastil_solubility_1982, kumar_modelling_1988, mendez-santiago_solubility_2012, tabernero_use_2011} and (iii) solvation models, parametrised with two key descriptors, an interaction energy term and an effective molecular volume.~\cite{cheng_calculation_2003, iwai_correlation_1992, shin_development_2001} Overall, these approaches are capable of reproducing solubilities over a broad range of pressures and temperatures with relative errors on the order of ~10\%. They exhibit however limited transferability and must be re-parametrised for each new solute and for any modification of the system (\textit{e.g.}, the presence of an interface or a cosolvent). \par

Conversely, molecular dynamics (MD) simulations are extensively employed to investigate CO$_2$ fluids~\cite{stubbs_molecular_2016}, from classical MD based on simple point-charge force fields~\cite{harris_carbon_1995, potoff_vaporliquid_2001, su_simulations_2006} to on-the-fly \textit{ab initio} MD.~\cite{mi_ab_2019, saharay_ab_2004} These methods have been applied to characterise the properties of CO$_2$ fluids in diverse settings, including the bulk phase,~\cite{mi_ab_2019, saharay_ab_2004} solvation processes,~\cite{noroozi_solvation_2016, noroozi_microscopic_2017, reddy_solubility_2019, su_simulations_2006} multicomponent mixtures,~\cite{idrissi_local_2010, potoff_vaporliquid_2001} and confined environments.~\cite{fuentes-azcatl_carbon_2019, rebiscoul_impact_2019} In particular, MD simulations enable the accurate determination of solvation free energies (typically on the order of a few kJ·mol$^{-1}$).~\cite{noroozi_solvation_2016, noroozi_microscopic_2017, stubbs_partial_2005, su_simulations_2006} They have significantly contributed to elucidating several key aspects of solvation in supercritical CO$_2$, including local density enhancements,~\cite{idrissi_local_2010, su_simulations_2006, yoon_molecular_2017} pronounced density inhomogeneities, and CO$_2$-philicity---that is, the specific affinity between CO$_2$ and particular chemical functionalities (for example, carbonyl groups).~\cite{altarsha_new_2012, azofra_theoretical_2013, ingrosso_electronic_2018, san-fabian_theoretical_2014} MD simulations offer multiple advantages: high predictive accuracy, direct access to microscopic scales, and good transferability across different molecular systems and thermodynamic state points. Their principal drawback, however, remains the computational cost. While the cost is relatively modest for analysing the solvation structure of a single solute (typically on the order of a few CPU·hours), it increases by roughly an order of magnitude for the calculation of thermodynamic properties, which requires the deployment of advanced sampling strategies (\textit{e.g.}, Widom insertion, thermodynamic integration, and related free-energy methods).~\cite{frenkel_understanding_2002} This computational overhead severely limits the routine application of molecular simulations to the modelling of CO$_2$ fluid properties in large-scale parametric studies covering extensive regions of thermodynamic phase space.\par

Classical density functional theory (cDFT) offers the best of both worlds.~\cite{mermin_thermal_1965, evans_nature_1979, lowen_density_2002, wu_density_2006, evans_new_2016, hansen_theory_2013}  The recent rise of the machine learning technologies provides a novel avenue for the construction of highly accurate excess functionals.~\cite{wu_perfecting_2023, simon_machine_2026} The central concept is to infer the functional dependence between the derivatives of the free energy and the density.~\cite{shang-chun_classical_2019, sammuller_neural_2023} This approach was initially formulated for one-dimensional systems, but it is now a rapidly evolving research area, with recent extensions to anisotropic particles (including CO$_2$)~\cite{simon_machine_2024, simon_orientational_2025, bui_unified_2026} and to ionic liquids~\cite{bui_learning_2025}, though it is restricted to external potentials with a planar geometry. A key outstanding issue concerns the transferability of these machine-learned functionals to external potentials that differ substantially from those represented in the training data, or to other thermodynamic conditions. The latter limit was recently addressed by a study which has demonstrated that training the functional exclusively in a region of the supercritical phase diagram is sufficient to predict the vapour–liquid coexistence curve and interfacial density profiles in the subcritical regime.~\cite{robitschko_learning_2025} These findings are promising for the future development of a machine-learned functional tailored to supercritical CO$_2$. \par

In this work, we follow a different route. We assume that the inhomogeneous excess free-energy fluctuations remain close to those of the corresponding homogeneous system, an approximation known as the homogeneous reference fluid approximation (HRF).~\cite{jeanmairet_molecular_2013-1} Within this framework, the excess free-energy functional is approximated by a quadratic functional that depends exclusively on the direct pair (two-body) correlation functions (DCF) of the pure solvent. 
This formalism has recently been combined with a molecular one-body density to construct a molecular DFT (MDFT) for water under ambient conditions.~\cite{zhao_molecular_2011,  jeanmairet_molecular_2013,  jeanmairet_molecular_2016, ding_efficient_2017} The spatial and orientational DCF were obtained from MD simulations of bulk water,~\cite{puibasset_bridge_2012, belloni_exact_2017} and the MDFT obtained qualitative successes for the calculation of the properties of structure and thermodynamics of the hydration of simple solutes, interfaces and biomolecules.~\cite{levesque_solvation_2012,  jeanmairet_hydration_2014, jeanmairet_study_2019, luukkonen_hydration_2020}  
To go beyond the limitations of the HRF, an additional corrective contribution, the bridge functional, must be introduced. This term can be evaluated with high accuracy using the weighted density approximation,~\cite{nordholm_generalized_1980, curtin_weighted-density-functional_1985, tarazona_free-energy_1985, tarazona_erratum_1985} albeit at a substantial numerical and technical cost. A more tractable strategy was subsequently proposed, in which the bridge functional is approximated as a cubic or quadratic functional of a coarse-grained density, leading to the coarse-grained bridge functional (CGB). The latter is characterised by only a few adjustable parameters (the coarse-graining length and the bulk equation of state).~\cite{borgis_simple_2020, borgis_accurate_2021} 
More recently, Bui and Cox introduced an alternative methodology that employs an inhomogeneous coarse-grained fluid density as a reference state.~\cite{bui_classical_2024, lum_hydrophobicity_1999} This coarse-grained density captures large-scale solvent fluctuations, such as those associated with surface tension, and is typically described by an excess free-energy functional expressed in a macroscopic form.~\cite{ten_wolde_drying-induced_2002} \par

On the basis of these results, we decided to assess the suitability of MDFT for modelling the solvation properties in scCO$_2$. MDFT requires three inputs, the CO$_2$–solute interaction potentials, the bulk two-body DCF and the parameters defining the bridge functional. In our previous work,~\cite{houssein_mohamed_molecular_2025} we determined the bulk two-body DCF for scCO$_2$ in the near-critical region using both MD simulations and an integral equation theory. 
In the present study, we construct and parametrise two different bridge functionals for scCO$_2$ and systematically compare MDFT predictions (structural properties and solvation free energies) with reference MD results. We determine the MDFT functional at a thermodynamic condition close to the critical point ($n_0 = 0.8~n_c$ and $T = 1.05~T_c$), and we apply the resulting framework to a set of solutes encompassing a broad range of chemical functionalities (water, CO$_2$, ethylene glycol, toluene, benzene, ethanol, methanol, isopropanol, ethane). The MDFT calculations exhibit excellent agreement with the MD benchmarks, demonstrating the capability of MDFT to provide robust and computationally efficient modelling of solvation properties in scCO$_2$. We note that the objective of the present work is not yet to establish transferability across the phase diagram but rather to assess whether MDFT can quantitatively reproduce solvation thermodynamics at a representative near-critical state point. In the longer term, MDFT-based modelling could facilitate the rational design and optimization of scCO$_2$-based processes and applications.  \par

The paper is structured as follows. In Section~\ref{sec:theory}, we introduce the MDFT formalism and the associated bridge functionals. The parameters employed in the MDFT calculations and MD simulations are described in detail in Section~\ref{sec:simdetails}. In Section~\ref{sec:results}, we present the development of the MDFT excess functional and subsequently discuss the comparison between MDFT predictions and MD reference data for a series of argon-like solutes and a set of molecular solutes, considering both structural properties and solvation free energies. Finally, the discussion is concluded in Section~\ref{sec:conclusion}.


\section{Theory}
\label{sec:theory}


\begin{figure}
\begin{center}
\includegraphics[width=0.7\columnwidth]{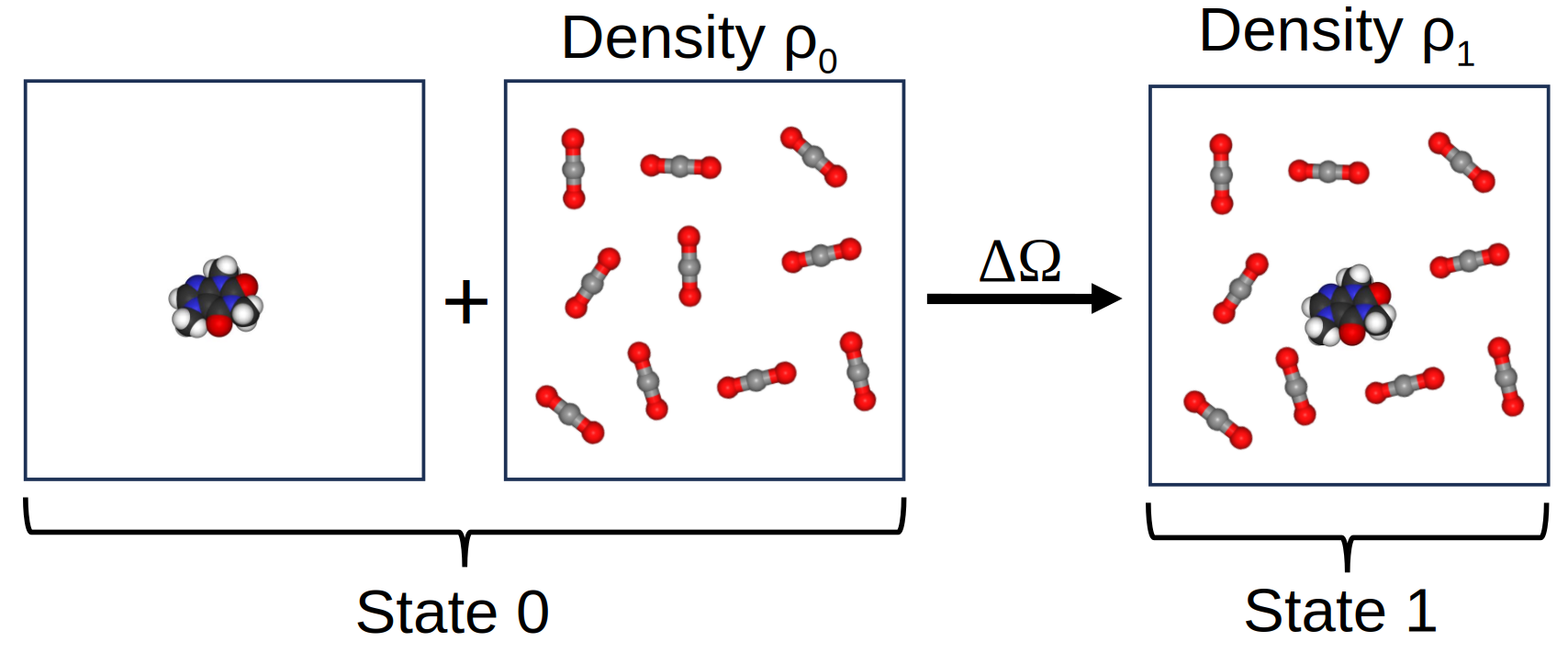}
\caption{Schematic representation of the solvation process investigated in this study. The initial state (labelled 0) consists of a solute in vacuum and a homogeneous bulk CO$_2$ characterized by the uniform density $\rho_0$. In the final state (labelled 1), the solute is fully solvated in CO$_2$, which is described by the equilibrium molecular density $\rho_1(\rr, \oo)$.}
\label{fig:solvation}
\end{center}
\end{figure}

MDFT addresses the thermodynamic solvation process illustrated in Figure~\ref{fig:solvation}. Both the solvation free energy, $\Delta \Omega$, and the equilibrium solvation structure, $\rho_1$, are determined by minimizing the solvation functional

\begin{align}
    \Delta \Omega[\rho] &= \Omega[\rho] - \Omega[\rho_0] \nonumber\\
    &= \Delta \Omega_\mathrm{id}[\rho]  + \Delta \Omega_\mathrm{exc}[\rho]  + F_\mathrm{ext}[\rho] 
    \label{eq:delta_omega_mdft}
\end{align}
where $\Omega$ is the grand potential of classical DFT, a functional of the solvent density $\rho$ and where $\rho_0 = n_0 / N_\omega$ is the uniform (bulk) solvent density, with $n_0$ the number density and $N_\omega=8\pi^2$ the angular normalization factor. The solvation free energy is decomposed in three contributions: $\Delta \Omega_\mathrm{id}$ is the ideal fluid contribution, 

\begin{equation}
    \Delta \Omega_\mathrm{id}[\rho] = k_BT \int \mathrm{d}1 \left[ \rho(1)\ln\left(\frac{\rho(1)}{\rho_0}\right) - \Delta \rho(1) \right]
    \label{eq:mdft_id}
\end{equation}
where $k_B$ denotes the Boltzmann constant, $T$ the absolute temperature, and the label $1$ stands for both the spatial coordinate $\rr_1$ and the three Euler angles $\oo_1 = (\theta_1, \phi_1, \psi_1)$ that specify the orientation of the solvent molecules, and $\Delta\rho = \rho - \rho_0$ the excess solvent density in the vicinity of the solute. The last term in eq~\ref{eq:delta_omega_mdft} corresponds to the contribution arising from the solvent–solute interaction,

\begin{align}
    F_\mathrm{ext} = \int\mathrm{d}1 \rho(1)v_\mathrm{ext}(1)
    \label{eq:mdft_ex}
\end{align}
where $v_\mathrm{ext}$ is the external potential that accounts for solute–solvent interactions, typically obtained from standard force fields. Finally, the excess contribution, $\Delta \Omega_\mathrm{exc}$, corresponds to the solvent–solvent interaction term. Its exact functional form is actually unknown and must therefore be approximated. We recall in the section S1 a formal definition of the excess contribution $\Delta \Omega_\mathrm{exc}$, as well as the derivation of the solvation function (eq~\ref{eq:delta_omega_mdft}) from the grand potential functional. \par

For dense fluids, it is reasonable to assume that solvent density fluctuations in the vicinity of the solute are comparable to those in the homogeneous bulk phase (\textit{i.e.}, in the absence of the solute).~\cite{chandler_gaussian_1993, sergiievskyi_solvation_2017} Within the framework of cDFT, it leads to the homogeneous reference fluid (HRF) functional, the quadratic Taylor expansion of the excess functional around the uniform bulk density $\rho_0$,

\begin{align}
    \Delta \Omega_\mathrm{HRF}[\rho] &= -\frac{k_BT}{2} \int\mathrm{d}1\mathrm{d}2 \Delta\rho(1) c(12) \Delta\rho(2) 
    \label{eq:mdft_hrf}
\end{align}
where $c(12)$ are the direct correlation functions (DCF) of the homogeneous bulk solvent (in the absence of the solute), which can be calculated from MD simulations or from integral equation theories.~\cite{ramirez_density_2002, puibasset_bridge_2012, belloni_exact_2017, houssein_mohamed_molecular_2025} The remaining contribution to the excess functional is referred to as the bridge functional, $\Delta \Omega_\mathrm{bridge} = \Delta \Omega_\mathrm{exc} - \Delta \Omega_\mathrm{HRF}$, and encompasses all solute-induced modifications of the solvent fluctuations. \par

In the following, we examine three approximations for the bridge functional. (1) The first consists in simply neglecting it, \textit{i.e.}, $\Delta \Omega_\mathrm{exc} \approx \Delta \Omega_\mathrm{HRF}$. This approximation has yielded qualitatively satisfactory results in studies of the hydration structure of simple solutes, interfaces, and biomolecules.~\cite{levesque_solvation_2012,  jeanmairet_hydration_2014, jeanmairet_study_2019, luukkonen_hydration_2020,  borgis_accurate_2021} It fails however to correctly predict the solvation free energy of macroscopic solutes because truncation at quadratic order is insufficient to capture the genuine density fluctuations of water, in particular the surface tension and the proximity to phase coexistence.~\cite{lum_hydrophobicity_1999, ten_wolde_drying-induced_2002, jeanmairet_molecular_2013-1, jeanmairet_molecular_2015, borgis_simple_2020} Some of us have attempted to introduce a posteriori correction to the solvation free energy based on the pressure,~\cite{jeanmairet_molecular_2015, sergiievskyi_fast_2014, sergiievskyi_solvation_2015} yet such correction does not modify the equilibrium density $\rho$ and remains ambiguous to define rigorously for a real solute. (2) The bridge functional can alternatively be approximated by employing a hard-sphere excess functional, \textit{e.g.}, the FMT excess functional.~\cite{rosenfeld_theory_1979, rosenfeld_free_1993, kahl_structure_1996, zhao_new_2011, zhao_correction_2011, liu_high-throughput_2013, liu_site_2013, li_multiscale_2020}. We detailed in the section S2.1 how to construct the hard-sphere function from the FMT excess function. It depends solely on a single parameter, the hard-sphere radius of the solvent. This parameter can be calibrated by enforcing agreement with the bulk free energy. (3) The final approximation considered is the coarse-grained bridge (CGB) functional. Since the microscopic interactions between solvent molecules are already accounted for by the HRF functional $\Delta \Omega^\mathrm{HRF}$, the bridge contribution can be modelled as a functional of a coarse-grained number density field $\bar{n}$,

\begin{align}
    \bar{n}(\rr) &= \int\dd\rr'n(\rr') \frac{1}{\sqrt{2\pi\sigma_g^2}^3} \exp\left( - \frac{\vert \rr - \rr'\vert}{2 \sigma_g^2} \right) \;,
    \label{eq:weight}
\end{align}
where $\sigma_g$ is the coarse-graining length and $n(\rr) = \int\dd\oo \rho(\rr, \omega)$ is the number density. For the coarse-grained number density $\bar{n}$, we assume that the bridge contribution is given as a sum of local contribution, obtained from the homogeneous equation of state. It reads

\begin{align}
    \Delta \Omega^\mathrm{CGB}_\mathrm{bridge}[\rho] = \int\dd\rr \Delta\omega^\mathrm{hom}_\mathrm{bridge}(\bar{n}(\rr)) \;,
    \label{eq:mdft_cgb}
\end{align}
where $\Delta\omega^\mathrm{hom}_\mathrm{bridge}(\rho)$ is the bridge solvation free energy per unit volume of the bulk fluid. This can be calculated exactly from the bulk equation of state, $P(n)$, when it is known (as explained in the section S2.2. Otherwise, $\Delta\omega^\mathrm{hom}_\mathrm{bridge}(n)$ can be approximated by a polynomial expansion in $\Delta n$ which is, by construction, at least of order 3, $\Delta\omega^\mathrm{hom}_\mathrm{bridge}(n) = a_3 \Delta n^3 + ... $. 
The parameters could be adjusted to some known properties of the bulk fluid. Eventually, the bridge functional $\Delta \Omega^\mathrm{CGB}_\mathrm{bridge}[\rho]$ (eqs~\ref{eq:weight}-\ref{eq:mdft_cgb}) depends on the coarse-graining length $\sigma_g$ and the polynomial expansion.

\section{Numerical Details}
\label{sec:simdetails}


In this study, we will compare MDFT and MD predictions for a series of solutes under near-critical thermodynamic conditions. We select the thermodynamic condition $0.8~n_0$ and $T = 1.05\,T_c$, and we investigate two classes of solutes: (i) spherical, argon-like species with different size and energy parameters, and (ii) nine molecular solutes---toluene, ethylene glycol, benzene, isopropanol, ethanol, methanol, water, CO$_2$, and ethane.
Among the existing CO$_2$ force fields~\cite{harris_carbon_1995, potoff_vaporliquid_2001, zhang_optimized_2005, cygan_molecular_2012}, we adopt the rigid EPM2 model proposed by Harris and Yung~\cite{harris_carbon_1995}. This model provides an accurate representation of the CO$_2$ phase diagram in the near-critical region while retaining a relatively simple functional form for the intermolecular interactions. In the EPM2 description, CO$_2$ is treated as a linear molecule with fixed C–O bond lengths, $d_{\mathrm{CO}} = 1.149$~\AA. Intermolecular interactions are represented as a sum of Lennard-Jones and Coulomb pair potentials. The corresponding critical parameters of this model are $T_c^\mathrm{EPM2} = 313.4$~K, $n_c^\mathrm{EPM2} = 10.3$~mol·L$^{-1}$, and $P_c^\mathrm{EPM2} = 76.5$~bar.
The argon-like solutes are modelled as uncharged Lennard-Jones sites with various Lennard-Jones length and well-depth parameters. For the molecular solutes, we employ the OPLS-AA force field,~\cite{jorgensen_development_1996} except for water, which is represented by the SPC/E model.~\cite{berendsen_missing_1987} The Lennard-Jones parameters for solute–CO$_2$ cross interactions are determined using the Lorentz–Berthelot combining rules. \par

MDFT calculations consist in minimizing the functional defined in eq~\ref{eq:delta_omega_mdft}, in which the ideal contribution is given by eq~\ref{eq:mdft_id}, the external contribution by eq~\ref{eq:mdft_ex}, and the excess contribution by the sum of the HRF functional (eq~\ref{eq:mdft_hrf}) and a bridge functional. We considered three types of bridge contributions: the absence of a bridge functional ($\Delta \Omega_\mathrm{bridge}[\rho]=0$), a coarse-grained bridge functional (CGB, eq~\ref{eq:mdft_cgb}), and a hard-sphere bridge functional (HSB). 
In MDFT, the main computational challenge arises from the six-dimensional dependence of the one-particle density (three spatial coordinates and three Euler angles). In previous work, some of us developed a highly efficient implementation of MDFT.~\cite{ding_efficient_2017} In that approach, the minimization is carried out using a combination of Fourier transforms and expansions in a rotationally invariant basis, which enables rapid evaluation of the various contributions to the MDFT functional. The spatial degrees of freedom are discretized on a uniform Cartesian grid within a cubic simulation box. In section S3, we provide a description of the different bases employed for the rotational degrees of freedom, as well as an analysis of the convergence of the various observables with respect to the box size and spatial mesh resolution. All molecular density functional theory (MDFT) calculations reported in this work were performed in a cubic box of side length $L_\mathrm{box} = 30$~\AA\ with a spatial grid spacing of $\Delta r = 0.3$~\AA, and a maximal angular index of $n_\mathrm{max} = 4$. For these parameters, an MDFT calculation requires less than one minute on a laptop computer, corresponding to a computational cost of approximately 0.01 CPU·hours.


MDFT calculations will be compared with MD simulations. We employed MD simulations with 4000 CO$_2$ molecules and a single solute molecule within a cubic box of length $92.13$~\AA. Periodic boundary conditions were applied in all directions. Long-range electrostatic interactions were evaluated using the Ewald summation with a particle-particle particle-mesh solver, employing a real-space cut-off of $r_\mathrm{cut} = 17$~\AA~and a target accuracy of $10^{-5}$.~\cite{hockney_computer_1988} The Lennard–Jones (LJ) interactions were truncated at the same distance, $r_\mathrm{cut} = 17$~\AA, and standard mean-field tail corrections to the energy and pressure were applied to account for the neglected long-range contributions.~\cite{sun_compass_1998} All simulations were carried out using the LAMMPS software package.~\cite{thompson_lammps_2022} Following an equilibration period of 1 ns in the NVT ensemble at \(T = 320~\text{K}\), the trajectories were propagated for an additional 9 ns under identical thermodynamic conditions, with the exception of benzene and toluene, for which longer production trajectories were required. The equations of motion were integrated using the velocity-Verlet algorithm with a time step of 1 fs, in combination with a Nosé–Hoover-type thermostat.~\cite{parrinello_polymorphic_1981, martyna_constant_1994, shinoda_rapid_2004} For each solute, solvation free energies were computed using the Bennett acceptance ratio method~\cite{bennett_efficient_1976} based on two independent MD simulations: a ``deletion'' trajectory, in which the solute interacts with the CO$_2$ solvent, and an ``insertion'' trajectory, in which the solute is absent. For the deletion trajectory, at intervals of 1~ps the instantaneous energy difference between the full system (CO$_2$ plus solute) and the corresponding solvent-only system was evaluated. Conversely, for the insertion trajectory, at intervals of 1~ps the energy difference between the pure CO$_2$ system and a hypothetical system comprising CO$_2$ and an inserted solute was calculated; in this case, the solute is removed immediately after each virtual insertion. Each trajectory thus yields 9000 deletion (respectively, insertion) energy samples, from which the Bennet method is applied to estimate the solvation free energy and its associated statistical uncertainty~\cite{frenkel_understanding_2002}. In parallel, site–site pair distribution functions were accumulated on-the-fly by LAMMPS every 5~ps from the deletion trajectory. Overall, the evaluation of solvation properties with MD simulations requires approximately 10 hours of computation on 64 CPU cores for each solute (\textit{i.e.}, a total computational cost of 640 CPU·hours). For the largest solutes (benzene, toluene, isopropanol), we have checked that the BAR results are equal (within error bars) with respect to more expansive methods (multiple BAR and the recently developed H4D~\cite{belloni_non-equilibrium_2019}). \par

To determine the homogeneous bridge functions and parametrized the hard-sphere and coarse-grained bridge functionals, we computed the pressure of homogeneous CO$_2$ at a series of reduced densities: 0.2, 0.4, 0.6, 0.8, 0.98, 1.22, 1.42, 1.60, 1.82, and 1.99~$\rho_c$ and at T=320~K. For each state point, the simulation cell contained 4000 CO$_2$ molecules, except at 0.8~$\rho_c$, for which 13\,500 molecules were used in order to improve sampling in the vicinity of the critical point. The same simulation parameters as for the solute calculations were employed. Each system was equilibrated for 2~ns, after which the pressure was sampled every picosecond over an additional 6~ns production run.

\section{Results}
\label{sec:results}

\subsection{Construction of the Functional}

We first determined the excess solvation functional $\Delta \Omega_\mathrm{exc}$, which is decomposed into the sum of the HRF contribution (eq~\ref{eq:mdft_hrf}) and the bridge functional. The former depends solely on the DCF of the bulk CO$_2$, which we have already evaluated in our previous work.~\cite{houssein_mohamed_molecular_2025} By combining exact MD simulation data for $r < 45$~\AA\ with the hypernetted-chain integral approximation for $r > 45$~\AA, we inverted the Ornstein–Zernike equations and obtained the DCF $c(12)$.\par

\begin{figure}[!h]
    \centering
    \includegraphics[width=0.45\textwidth]{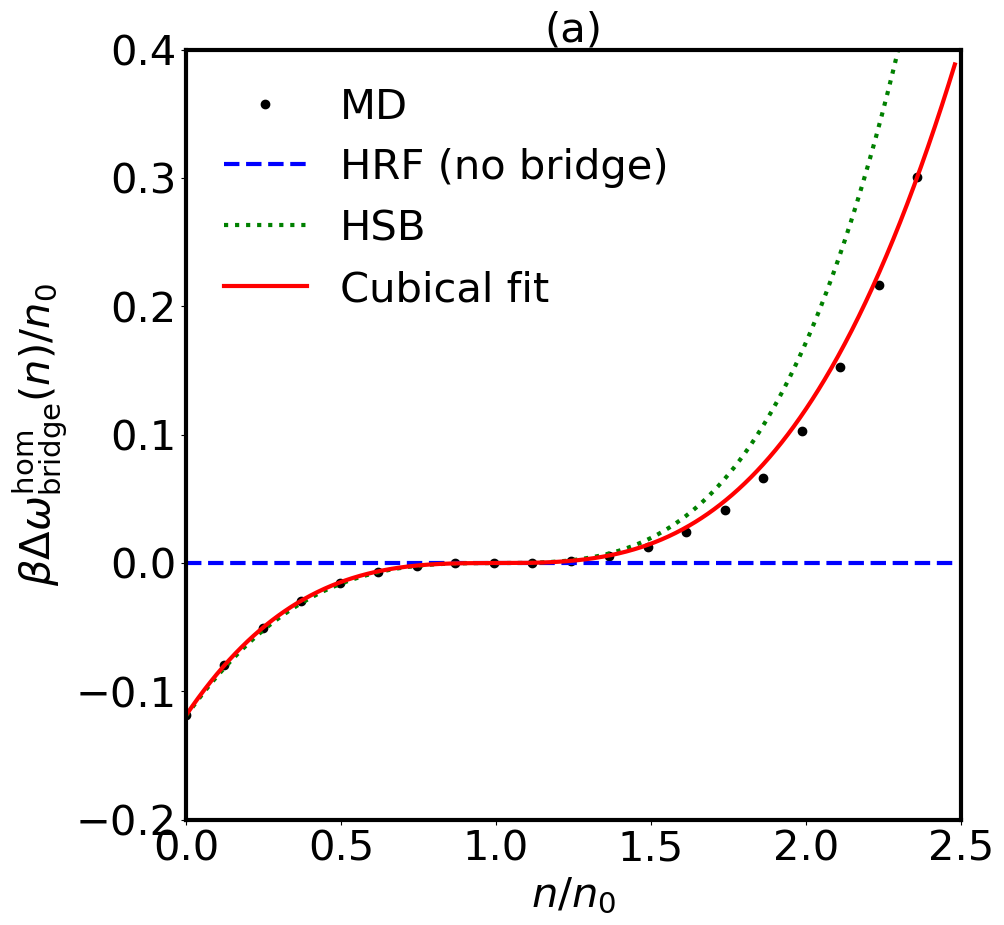}
    \includegraphics[width=0.42\textwidth]{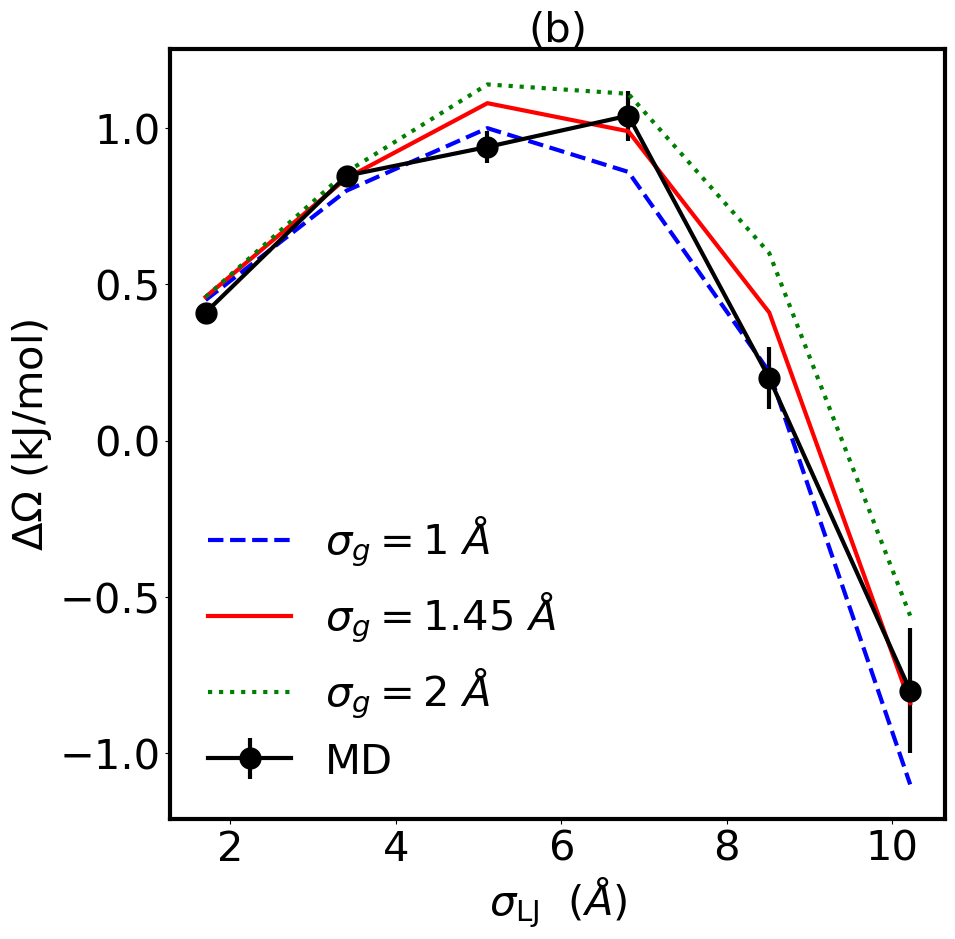}
    \caption{(a) Bridge functions for a bulk fluid with different levels of theory. MD: exact results obtained from the MD-based equation of state (symbols), no bridge (blue line), HSB: hard-sphere bridge functions with $r_\mathrm{HS} = 1.78$~\AA\ (green line), and  cubical fit, $a \Delta n^3$, with $a = 0.119~k_BT/n_0^2$ (red line). (b) Solvation free energy for a series of 6 Lennard–Jones solutes characterized by the interaction parameter $\epsilon_\mathrm{LJ} = 0.25$~kJ/mol and varying $\sigma_\mathrm{LJ}$. Discrete symbols represent MD simulation results, whereas continuous curves correspond to predictions from the CGB method obtained using three distinct coarse-graining length, $\sigma_g = 1$, $1.45$, and $2$~\AA, plotted in blue, red, and green, respectively. The coarse-graining length $\sigma_g = 1.45$~\AA\ reproduces well the MD results.}
    \label{fig:homogeneous_bridge}
\end{figure}

For the bridge functionals, we examined three distinct approximations in this study: the absence of a bridge functional (HRF), the hard-sphere functional (HSB), and the coarse-grained bridge functional (CGB). As a first step, we calibrated the parameters of both the hard-sphere and coarse-grained bridge functions to reproduce the homogeneous bridge contribution, which is evaluated exactly from the equation of state (see eq~S27). The performance of these approximations after parameter optimization, compared with the exact homogeneous bridge, is presented in Figure~\ref{fig:homogeneous_bridge}(a).
The hard-sphere bridge depends on a single parameter, namely the hard-sphere radius, for which we obtained $r_\mathrm{HS}=1.78$~\AA\ after fitting. For the coarse-grained bridge, we first observe that the homogeneous bridge function exhibits an approximately cubic dependence on the density deviation. We therefore approximated the coarse-grained homogeneous bridge function by a polynomial expansion of order 3, $\Delta \omega^\mathrm{hom}_\mathrm{bridge}(n) \simeq a_3\, \Delta n^3$, neglecting higher orders in the polynomial expansion. This implies that the coefficient $a_3$ is given by $ a_3 = -\Delta \omega^\mathrm{hom}_\mathrm{bridge}(0)/n_0^3$
The coarse-graining length, which enters the definition of the coarse-grained density through eq~\ref{eq:weight}, is finally determined by fitting the solvation free energies of a series of the most repulsive Lennard–Jones particles, with $\epsilon_\mathrm{LJ}=0.25$~kJ/mol and radii ranging from $1.7$~\AA to $10.2$~\AA. The corresponding results are reported in Figure~\ref{fig:homogeneous_bridge}(b) and show that the value $\sigma_g = 1.45$~\AA\ reproduce accurately the MD data.

\subsection{Solvation of spherical solutes}

\begin{figure}[!h]
    \centering
    \begin{subfigure}{0.45\textwidth}
         \includegraphics[width=1\textwidth]{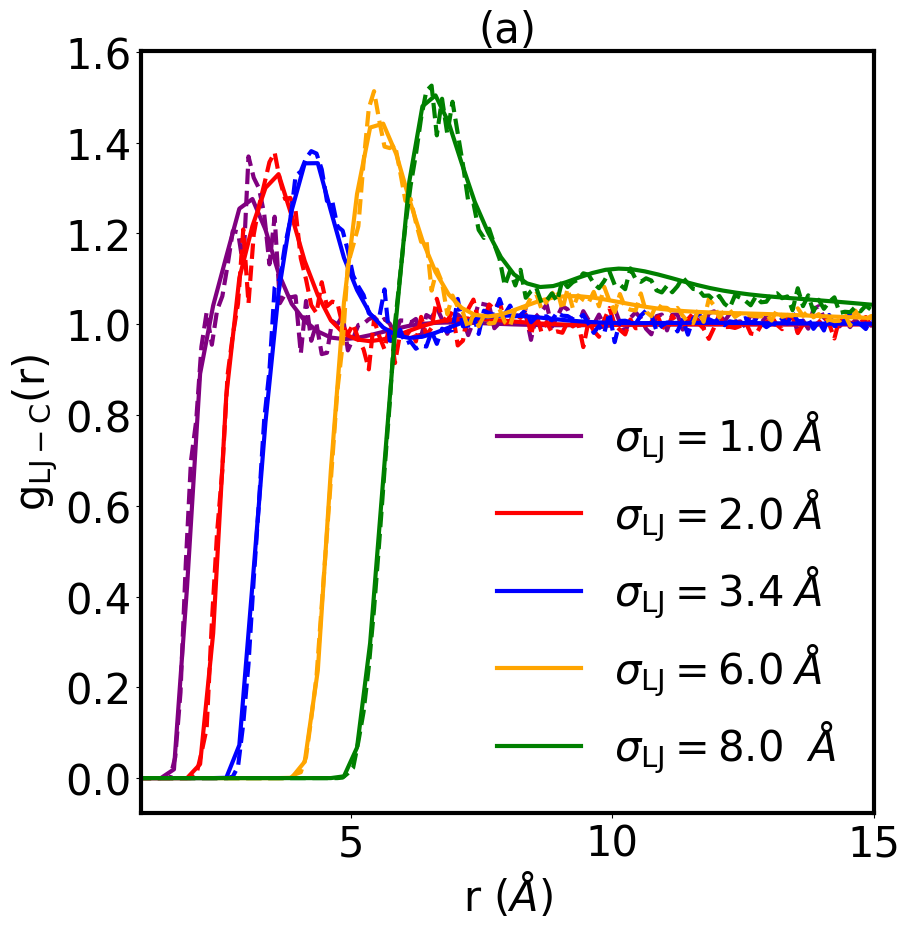}
     \end{subfigure}
     \begin{subfigure}{0.45\textwidth}
         \includegraphics[width=1\textwidth]{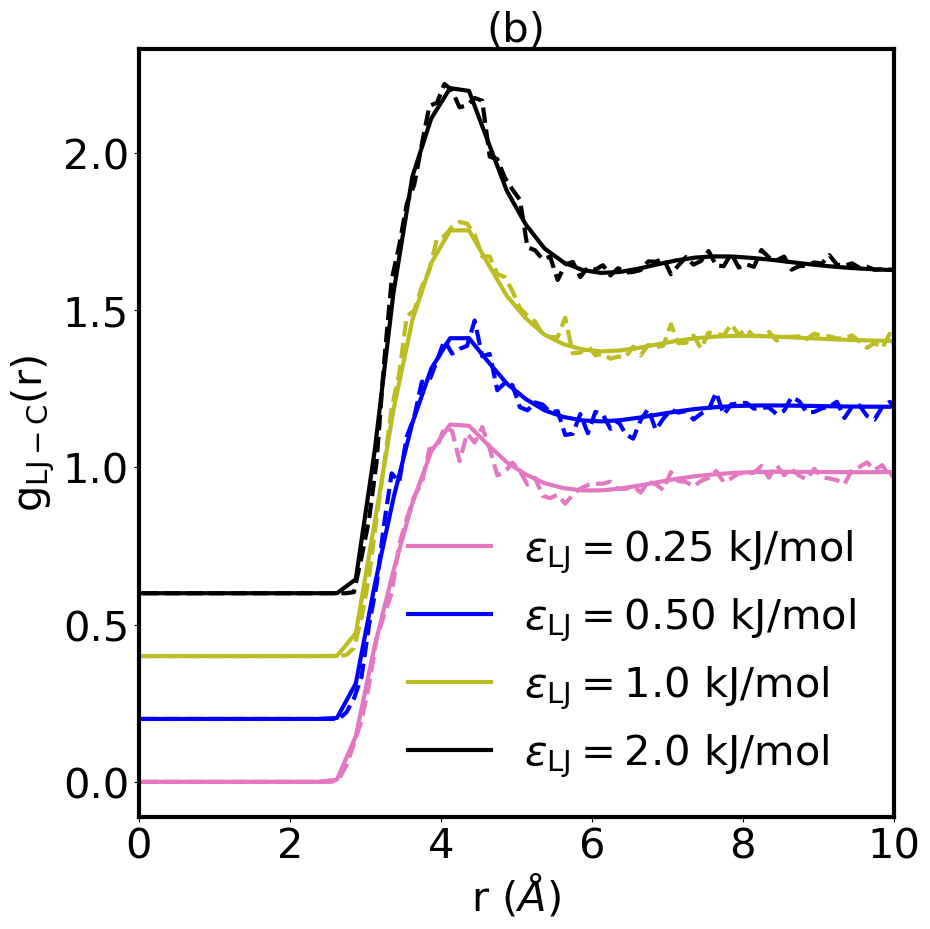}
     \end{subfigure}
\caption{Comparison of the solvation structure of supercritical CO$_2$ in the vicinity of a spherical solute obtained by MD and MDFT: pair distribution function between the LJ solute and the carbon atom of the CO$_2$ solvent molecules, computed using two approaches: MDFT with CGB functional predictions (solid line) and MD calculations (dashed line). We studied two sets of LJ solutes: (a) solutes with varying LJ radii $\sigma_\mathrm{LJ}$ and a fixed interaction energy $\epsilon_\mathrm{LJ} = 1$~kJ/mol, and (b) solutes with a fixed LJ radius $\sigma_\mathrm{LJ} = 3.405$~\AA~and different interaction energies $\epsilon_\mathrm{LJ}$. The pair distribution functions are shifted upward from the lowest $\epsilon_\mathrm{LJ}$ (0.25~kJ/mol) to the highest (2.0~kJ/mol).}
\label{fig:structure_lj}
\end{figure}

\begin{figure}[!h]
    \centering
    \includegraphics[width=0.8\textwidth]{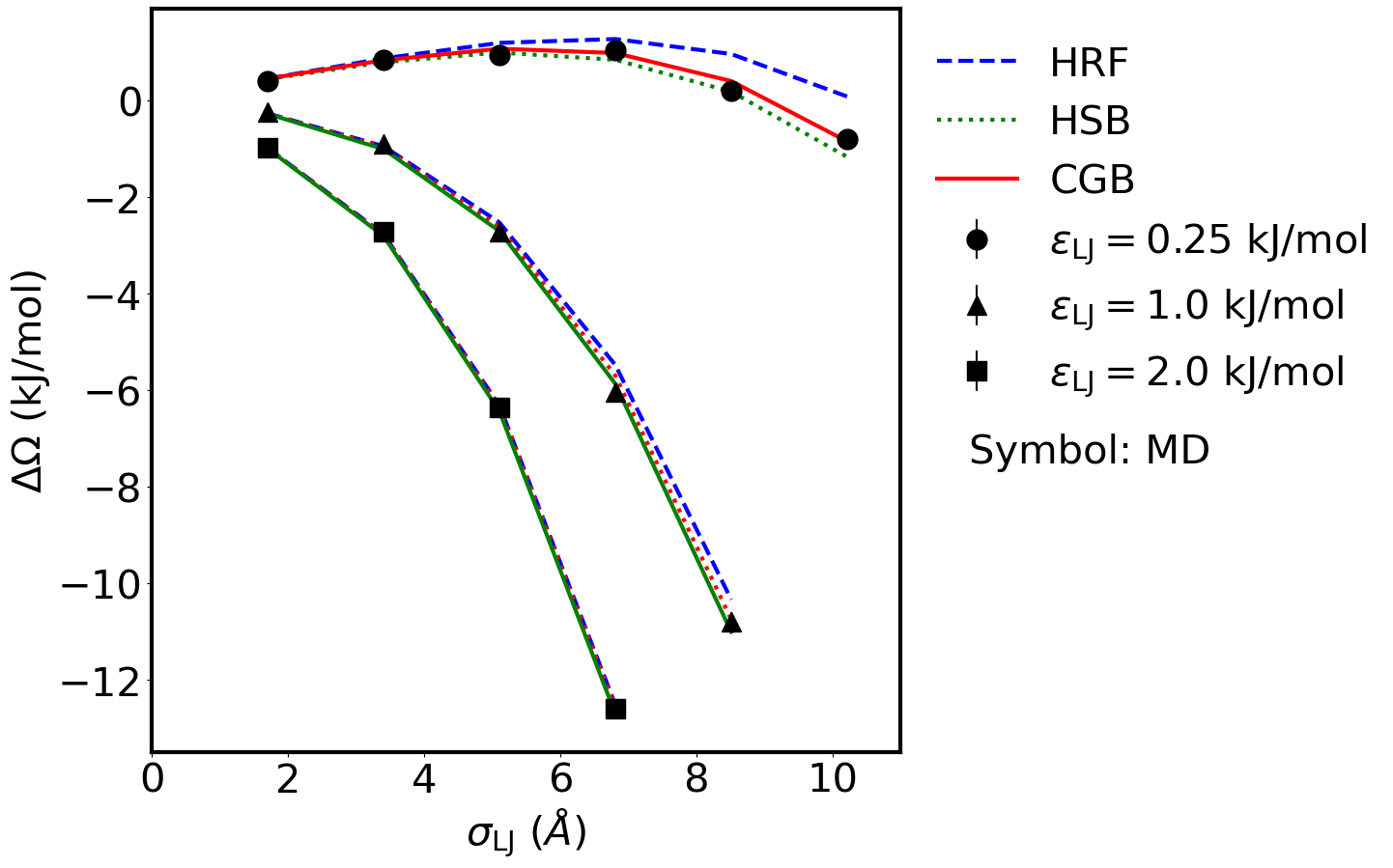}
    \caption{Comparison of the solvation free energies of spherical solutes in supercritical CO$_2$ obtained from MD and MDFT. The MD results are represented by distinct symbols corresponding to different solute–solvent interaction energies. The MDFT predictions were generated using three distinct approximations: the no-bridge functional approximation (HRF, blue line), the hard-sphere bridge functional (HSB, green line), and the coarse-grained bridge functional (CGB, red line).}
    \label{fig:SFE_lj}
\end{figure}

We first evaluated the accuracy of this new MDFT approach by comparing its structural and thermodynamic predictions with MD results for spherical solutes. To this end, we investigated four series of solutes characterized by Lennard-Jones interaction energies of $\epsilon_\mathrm{LJ} = 0.25, 0.5, 1.0,$ and $2.0$~kJ/mol and by interaction length parameters ranging from $1$ to $10$~\AA. MDFT predictions are identical for the different versions of the bridge functionals. Figure~\ref{fig:structure_lj} presents the radial distribution functions between the various LJ solutes and the carbon atom of the CO$_2$ molecule, $g_\mathrm{LJ-C}(r)$. For all solute radii and interaction energies considered, the MDFT predictions exhibit an essentially perfect overlap with the MD results, indicating an excellent agreement. 
We subsequently examined the solvation free energies. The MDFT results are reported in Figure~\ref{fig:SFE_lj}, together with three variants of MDFT: the no-bridge functional (HRF) approximation, the hard-sphere bridge functional (HSB), and the coarse-grained bridge functional (CGB). Overall, we again observe an excellent agreement between the solvation free energies obtained from MD and those from MDFT, with the exception of the smallest interaction energy ($\epsilon = 0.25$~kJ/mol) in combination with the largest solute sizes ($\sigma > 7$~\AA). For these solutes, the HRF approximation systematically overestimates the solvation free energies relative to the MD results, and the discrepancy increases with solute size. These solutes are similar to large cavities, a regime in which the no-bridge functional approximation is known to break down. In contrast, MDFT including bridge functional corrections (HSB and CGB) yields nearly indistinguishable solvation free energies, in close quantitative agreement with the MD data over a broad range of solute sizes (up to 10~\AA) and solvation free energies (from approximately $-13$ to $1$~kJ/mol). These observations motivated us to further evaluate the predictions of MDFT for molecular solutes.

\subsection{Solvation of molecular solutes}

\begin{figure}[!h]
    \centering
    \includegraphics[width=0.9\textwidth]{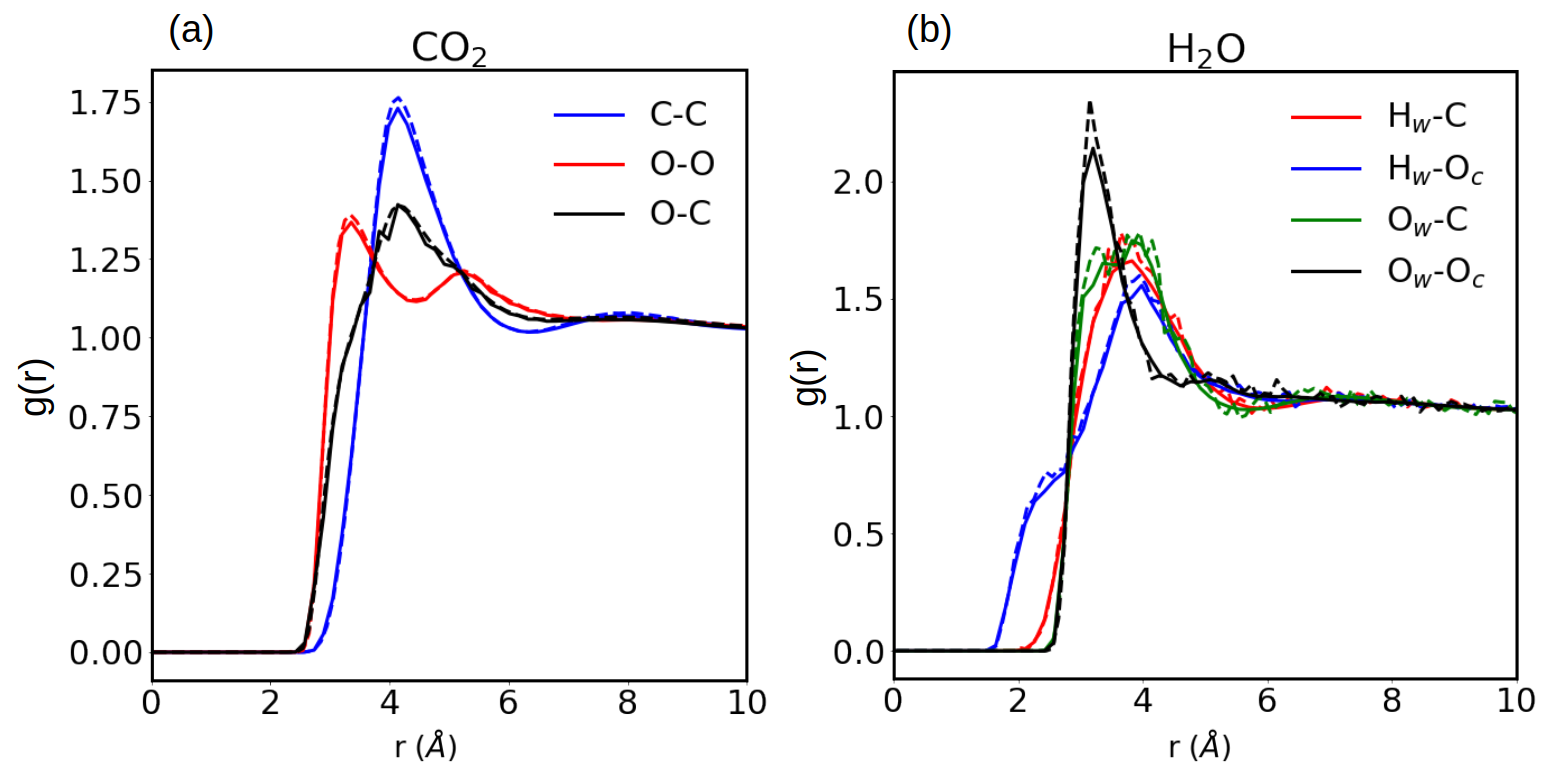}
    \caption{Comparison of the solvation structures surrounding a molecular solute: (a) CO$_2$ and (b) H$_2$O. All site–site pair distribution functions are presented, with the MDFT predictions represented by solid lines and MD simulation results indicated by dashed lines.}
    \label{fig:structure_mol}
\end{figure}

We next turned to a broader set of molecular solutes: toluene, ethylene glycol, benzene, isopropanol, ethanol, methanol, water, CO$_2$, and ethane. For each, we evaluated the site–site radial distribution functions (RDF) using both MD simulations and MDFT calculations. Figure~\ref{fig:structure_mol} presents the RDFs for water and CO$_2$ solutes, revealing an almost perfect overlap between MD and MDFT. RDF for the remaining solutes, shown in Figure~S3, likewise confirm that MDFT quantitatively captures the local solvation structure around a wide variety of molecular solutes. \par 

\begin{figure}[!h]
    \centering
    \includegraphics[width=0.8\textwidth]{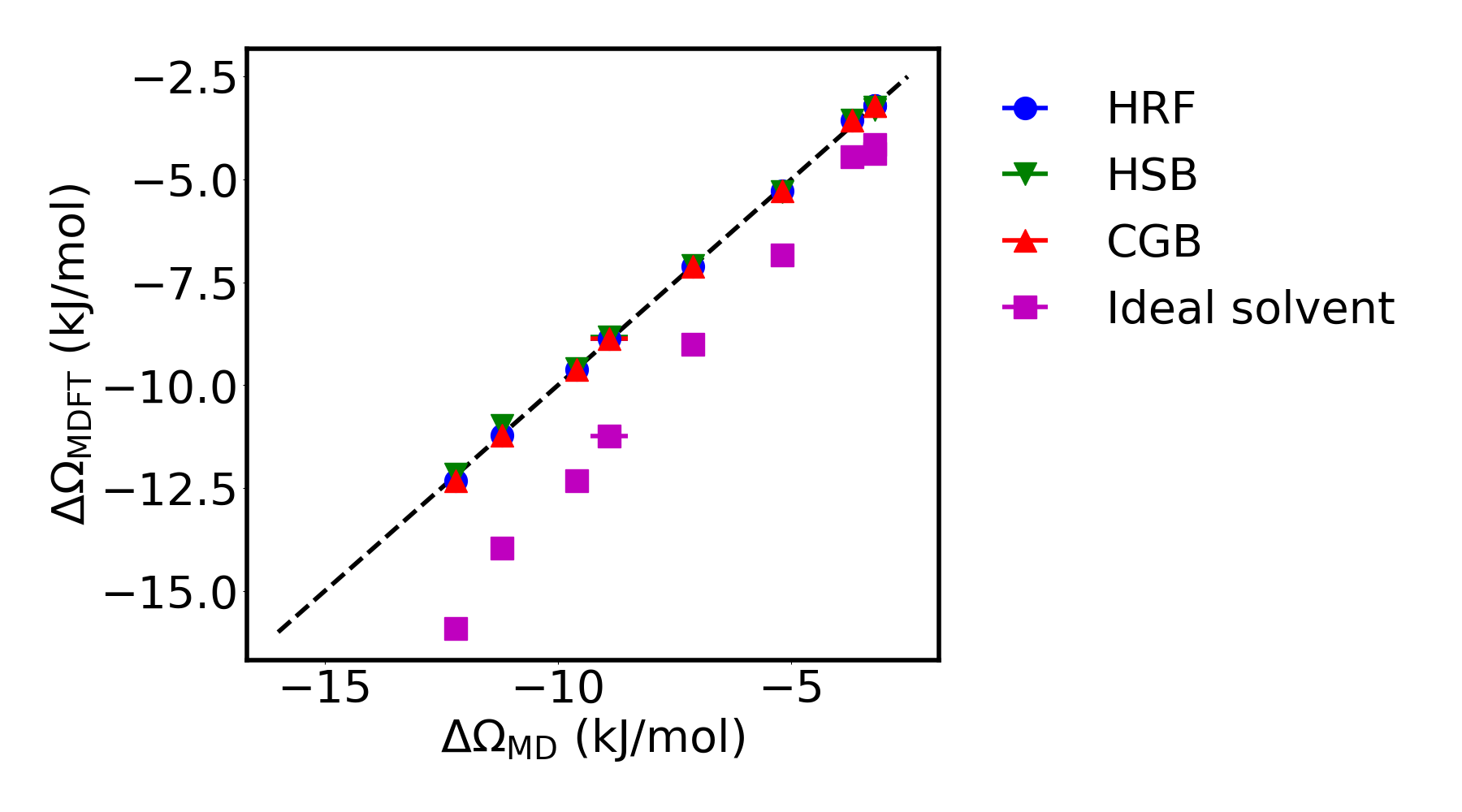}
    \caption{Correlation between the solvation free energies obtained from MD simulations and MDFT calculations for a series of molecular solutes: toluene, ethylene glycol, benzene, isopropanol, ethanol, methanol, water, CO$_2$, and ethane, ordered according to their MD solvation free energies. The MDFT estimates are computed using different approximations of the bridge functional (HRF, HSB, CGB), whereas the ideal solvent values correspond to MDFT calculations performed in the absence of an excess free-energy contribution (i.e., with $\Delta\Omega_\mathrm{exc}=0$). The dashed line represents the line of identity.}
    \label{fig:SFE_mols}
\end{figure}

Finally, we evaluated the solvation free energy predictions obtained from MDFT against those derived from MD simulations for the same set of solutes. Within the MDFT framework, we considered the version without a bridge functional (HRF), the two formulations including a bridge functional (HSB and CGB) and the case of an ideal solvent, in which only the ideal and external contributions are retained in the solvation functional (eq~\ref{eq:delta_omega_mdft}), \textit{i.e.}, $\Delta\Omega_\mathrm{exc}=0$. As shown in Figure~\ref{fig:SFE_mols}, all three MDFT variants reproduce the MD solvation free energies over a broad range (from -12 to -1 kJ/mol) with excellent accuracy, yielding an average relative error of approximately 1~\%. 
Moreover, the three MDFT variants yield indistinguishable results, indicating that the contribution of the bridge functional is negligible for the solutes investigated in this work.
The negligible contribution of the bridge functional contrasts with the situation encountered for liquid water previously investigated with MDFT. This difference arises from the absence, in scCO$_2$, of a strongly directional bond network. Consequently, density fluctuations in the vicinity of molecular solutes are already captured with high accuracy at the HRF level.
Finally, the solvation free energies obtained for the idealized solvent deviate from the MD reference values by few kJ/mol, with a mean relative error of approximately 30~\%. Beyond this excellent agreement, it is also pertinent to compare the computational cost of the two approaches. A single MDFT calculation requires only 0.01 CPU·hours, whereas the corresponding MD simulations require approximately 600 CPU·hours. These results demonstrate that the MDFT framework is capable of accurately reproducing both the solvation free energies and the microscopic structural properties of molecular solutes, while requiring roughly 60,000 times less computational effort.  \par

\section{Conclusion}
\label{sec:conclusion}

In this work, we demonstrate the effectiveness of molecular density functional theory (MDFT) in accurately reproducing both the solvation structure and the solvation free energy in supercritical CO$_2$ (scCO$_2$). Reliable modelling of these properties is crucial for the rational design and optimization of novel scCO$_2$-based industrial processes, for instance in the context of Green Chemistry, where scCO$_2$ serves as a highly efficient and environmentally benign solvent. We show here that MDFT calculations obtain the same results as MD simulations with near-quantitative accuracy (both in terms of radial distribution functions and solvation free energies) while incurring a computational cost that is 60,000 times lower. To achieve this level of accuracy, we constructed the MDFT functional based on three principal components: (i) the adoption of the homogeneous reference fluid (HRF) approximation, in which the excess functional is decomposed into an HRF contribution that depends on bulk solvent correlations and an additional correction term; (ii) the use of bulk solvent correlations derived from MD simulations to evaluate the HRF contribution; and (iii) the use of the molecular density formalism, wherein the solvent density is explicitly treated as a function of both spatial position and molecular orientation. We perform a comparative analysis of MD and MDFT for a series of argon-like solutes spanning different sizes and interaction strengths, as well as for a set of molecular solutes. Beyond the excellent overall agreement between the two approaches, our results indicate that the correction term to the HRF approximation is negligible. This observation provides a promising basis for extending this MDFT framework to a broader range of thermodynamic conditions. \par

The functional developed in this work is, at present, restricted to a single thermodynamic state. In the near future, we plan to assess its performance in more complex situations, such as in the presence of interfaces or under confinement—conditions that are particularly relevant for modelling CO$_2$ storage in geological reservoirs. To describe more realistic scenarios, however, it is essential to investigate how solvation and interfacial structures vary with temperature and pressure, and to design MDFT functionals that are transferable across the entire phase diagram. The emergence of machine-learning-based functionals, when combined with the molecular density functional formalism, appears to be a particularly promising direction for future developments.

\section*{Code availability}

The MDFT code is available on request from the corresponding author.

\section*{Acknowledgements}

The authors would like to thank the local HPC cluster at LPCT for computing time. High Performance Computing resources were partially provided by the EXPLOR centre hosted by the University de Lorraine (project 2019CPMXX1332). This project was provided with computing HPC and storage resources by GENCI at TGCC thanks to the grant 2022-AD010813681 on the supercomputer Joliot Curie’s ROME partition. M.H.M. gratefully acknowledges the University of Lorraine and the C2MP PhD school for a PhD grant. O.T., C.M. and A.C. acknowledge financial support from Agence Nationale de la Recherche through project BAC2MOL (ANR-24-CE29-4800-01).

\section*{Data availability}

The Jupyter notebook used to generate the figures and the inputs for the LAMMPS molecular dynamics simulations have been deposited in the dataverse of the Université de Lorraine at: \url{https://doi.org/10.57745/86IWUT}.

\newpage

\setcounter{section}{0}
\setcounter{figure}{0}
\setcounter{table}{0}
\setcounter{equation}{0}

\renewcommand{\thesection}{S\arabic{section}}
\renewcommand{\thefigure}{S\arabic{figure}}
\renewcommand{\thetable}{S\arabic{table}}
\renewcommand{\theequation}{S\arabic{equation}}

\makeatletter
\renewcommand{\@seccntformat}[1]{\csname the#1\endcsname\quad}
\makeatother

\section{Construction of the Functional}
\label{sec:si_construction}

\subsection{Notation in MDFT}

The grand potential functional within the framework of Molecular Density Functional Theory (MDFT) is given by

\begin{align}
    \Omega_{v_\mathrm{ext}}\left[ \rho \right] = \mathcal{F}[\rho] - \mu N[\rho] + F_\mathrm{ext}[\rho] \;,
\end{align}
where $\mathcal{F}$ is the intrinsic functional that does not depend on the external potential $v_\mathrm{ext}$, $\mu$ is the chemical potential, $N[\rho] = \int\dd\rr\dd\oo \rho(\rr,\omega)$, $F_\mathrm{ext}[\rho] = \int\dd\rr\dd\oo v_\mathrm{ext}(\rr,\oo) \rho(\rr,\oo)$ and the spatial ($\rr$) and angular ($\oo$) degrees of freedom have been made explicit. The intrinsic functional $\mathcal{F}[\rho]$ can be decomposed into two contributions: the ideal-gas functional $\mathcal{F}_\mathrm{id}$ and the excess part $\mathcal{F}_\mathrm{exc}$, the latter accounting for inter-particle interactions,

\begin{align}
\mathcal{F}[\rho]=\mathcal{F}_{\mathrm{id}}[\rho]+\mathcal{F}_{\mathrm{exc}}[\rho].
\end{align}
The ideal contribution (\textit{i.e.}, the case in the absence of interactions) can be expressed as

\begin{align}
\mathcal{F}_{\mathrm{id}}[\rho]=k_BT\int \dd\rr\dd\oo \rho(\rr,\oo)\left(\ln\left(\lambda^{3}\rho(\rr,\oo)\right)-1\right) \;.
\label{eq:f_id}
\end{align}
Using the excess contribution, the one- and two-body direct correlation functionals can be defined as functional derivatives of the excess potential,

\begin{align}
c^{(1)}(\rr,\oo; \rho)&=-\frac{\delta\beta\mathcal{F}_\mathrm{exc}}{\delta\rho(\rr,\oo)}[\rho] \label{eq:fonctions_c_1_corps_cdft}\\
c^{(2)}(\rr,\oo,\rr',\oo'; \rho)&=\frac{\delta c^{(1)}(\rr,\oo; \rho)}{\delta\rho(\rr',\oo')}=-\frac{\delta^{2}\beta\mathcal{F}_\mathrm{exc}}{\delta\rho(\rr',\oo')\delta\rho(\rr,\oo)}\\
&=c^{(2)}(\rr',\oo',\rr,\oo; \rho) \label{eq:fonctions_c_2_corps_cdft}
\end{align}
They are functions of the particle positions and orientations, $(\rr,\oo)$, and at the same time functionals of the one-body density, $\rho$. In particular, $c^{(1)}(\rho_0)$ is directly related to the excess chemical potential of the homogeneous bulk system,

\begin{align}
    c^{(1)}(\rho_0) = -\frac{\delta\beta\mathcal{F}_\mathrm{exc}}{\delta\rho(\rr,\oo)}[\rho=\rho_0] = -\beta\mu + \beta\mu_{id}(\rho_0) = -\beta \mu_\mathrm{exc} \;,
    \label{eq:bulk_potential}
\end{align}
where $\mu_{id}(\rho_0)=k_{B}T\ln\bigl(\lambda^{3}\rho_0\bigr)$ denotes the chemical potential of an ideal gas at the reference density $\rho_0$. 

\subsection{Solvation functional}

For a homogeneous bulk fluid, $v_\mathrm{ext} =0$, and the equilibrium density is $\rho_0$. In this case, the grand potential is given by

\begin{align}
    \Omega_{v_\mathrm{ext}=0}[\rho_0]=\mathcal{F}[\rho_0]-\mu N_0 \;,
\end{align}
where $N_0 = \int\dd\rr\dd\oo \rho_0$
The difference between the grand potential functional in the presence of an external potential and its bulk counterpart is given by

\begin{align}
\Delta \Omega[\rho] &= \Omega_{v_\mathrm{ext}}[\rho]-\Omega_{v_\mathrm{ext}=0}[\rho_0] \nonumber \\
&=\mathcal{F}[\rho]-\mathcal{F}[\rho_0]+ F_\mathrm{ext}[\rho] - \mu\Delta N[\rho] \\
&= (\mathcal{F}_\mathrm{id}[\rho]-\mathcal{F}_\mathrm{id}[\rho_0]) + (\mathcal{F}_\mathrm{exc}[\rho]-\mathcal{F}_\mathrm{exc}[\rho_0]) + F_\mathrm{ext}[\rho]-\mu\Delta N[\rho] \;.
\label{eq:exact_functional}
\end{align}
where $\Delta N[\rho] = N[\rho] - N_0$. The difference between the excess functionals can be obtained by performing a thermodynamic integration between the reference state (with density $\rho_0$, and the state of interest with density $\rho$. By employing eq~\ref{eq:fonctions_c_1_corps_cdft} and eq~\ref{eq:fonctions_c_2_corps_cdft}, this difference can be expressed as

\begin{align}
    (\mathcal{F}_\mathrm{exc}[\rho]-\mathcal{F}_\mathrm{exc}[\rho_0]) &= -k_BT\int_0^1\dd\lambda\int\dd\rr\dd\oo \Delta \rho(\rr,\omega)c^{(1)}(\rr,\oo; \rho_\lambda) \\
    &= -k_BT \int\dd\rr\dd\oo\int\dd\rr'\dd\oo' \Delta \rho(\rr,\omega) \mathcal{C}(\rr,\oo, \rr', \oo') \Delta \rho(\rr',\omega') \nonumber \\
    &-k_BT \Delta N[\rho] c^{(1)}(\rho_0) 
    \label{eq:thermo_int_exc}
\end{align}
where

\begin{align}
    \mathcal{C}(\rr,\oo,\rr',\oo') = \int^{1}_{0}d\alpha(\alpha-1) c^{(2)}(\rr,\oo,\rr',\oo'; \rho_\alpha) \;,
\end{align}
$\Delta \rho = \rho - \rho_0$. Finally, the solvation free energy can be reformulated as the sum of three distinct contributions,

\begin{align}
    \Delta \Omega[\rho] &= \underbrace{(\mathcal{F}_\mathrm{id}[\rho]-\mathcal{F}_\mathrm{id}[\rho_0])
    -k_BT \Delta N[\rho] c^{(1)}(\rho_0) -\mu \Delta N[\rho]}_{\Delta\Omega_\mathrm{id}[\rho]} \nonumber\\
    &  \underbrace{-k_BT \int\dd\rr\dd\oo\int\dd\rr'\dd\oo' \Delta \rho(\rr,\omega) \mathcal{C}(\rr,\oo, \rr', \oo') \Delta \rho(\rr',\omega')}_{\Delta\Omega_\mathrm{exc}[\rho]} 
    + F_\mathrm{ext}[\rho] \\
    &= \Delta \Omega_\mathrm{id}[\rho] + \Delta \Omega_\mathrm{exc}[\rho] + F_\mathrm{ext}[\rho]
\end{align}
 the relationship between \(c^{(1)}(\rho_0)\) and the excess chemical potential (eq~\ref{eq:bulk_potential}), the expression for the ideal solvation functional can be recast in a simplified form.

\begin{align}
    \Delta \Omega_\mathrm{id}[\rho] &= k_BT\int \Big[\rho(\rr,\oo)\ln{\Big(\frac{\rho(\rr,\oo)}{\rho_0}\Big)}-\rho(\rr,\oo)+\rho_0\Big]\dd\rr\dd\oo \label{eq:fonct_id}
\end{align}
while the excess solvation functional reads

\begin{align}
    \Delta \Omega_\mathrm{exc}[\rho] &= \mathcal{F}_\mathrm{exc}[\rho] - \mathcal{F}_\mathrm{exc}[\rho_0] + k_BT \Delta N[\rho] c^{(1)}(\rho_0) \\
    &=  \mathcal{F}_\mathrm{exc}[\rho] - \mathcal{F}_\mathrm{exc}[\rho_0] -\frac{\delta\mathcal{F}_\mathrm{exc}}{\delta\rho(\rr,\oo)}[\rho=\rho_0] \Delta N[\rho]
    \label{eq:delta_omega_exc}
\end{align}

\section{Bridge functionals}

Within the MDFT framework, the excess solvation functional is partitioned into two distinct contributions: the HRF functional, which accounts for fluctuations in the homogeneous fluid, and the so‑called bridge functional, which incorporates all higher‑order corrections. It reads

\begin{align}
    \Delta \Omega_\mathrm{exc}[\rho] =  \Delta \Omega_\mathrm{HRF}[\rho] +  \Delta \Omega_\mathrm{bridge}[\rho]
    \label{eq:mdft_exc_decompo}
\end{align}
where

\begin{align}
    \Delta \Omega_\mathrm{HRF}[\rho] &= -\frac{k_BT}{2} \int\dd\rr\dd\oo\int\dd\rr'\dd\oo' \Delta\rho(\rr,\oo) c(\rr,\oo,\rr',\oo') \Delta\rho(\rr',\oo') \;,
    \label{eq:mdft_hrf_full}
\end{align}
where $c(\rr,\oo,\rr',\oo') = c^{(2)}(\rr,\oo,\rr',\oo'; \rho_0)$ is the direct correlation function (DCF) of the homogeneous bulk system. In this work, we assess three different approximations of the bridge functional: (1) the HRF approximation ($\Delta \Omega_\mathrm{bridge}[\rho] = 0$); (2) the hard-sphere bridge approximation; and (3) the coarse-grained bridge approximation. Using eq~\ref{eq:mdft_exc_decompo} and eq~\ref{eq:delta_omega_exc}, the bridge functional can be related to the excess functional as

\begin{align}
      \Delta \Omega_\mathrm{bridge}[\rho] &= \Delta \Omega_\mathrm{exc}[\rho] -  \Delta \Omega_\mathrm{HRF}[\rho] \\
     &= \mathcal{F}_\mathrm{exc}[\rho] - \mathcal{F}_\mathrm{exc}[\rho_0] -\frac{\delta\mathcal{F}_\mathrm{exc}}{\delta\rho(\rr,\oo)}[\rho=\rho_0] \Delta N[\rho] -  \Delta \Omega_\mathrm{HRF}[\rho]
    \label{eq:mdft_bridge_decompo}
\end{align}

\subsection{Hard-sphere bridge (HSB)}

For a hard-sphere fluid, fundamental measure theory (FMT) yields a highly accurate representation of the excess free-energy functional, $\mathcal{F}_\mathrm{exc}^\mathrm{FMT}$.~\cite{rosenfeld_free-energy_1989, roth_fundamental_2010} The hard-sphere bridge strategy is based on approximating the bridge functional by its FMT counterpart, \textit{i.e.}, $\Delta \Omega_\mathrm{bridge} \simeq \Delta \Omega_\mathrm{bridge}^\mathrm{FMT}$. The FMT bridge functional depends on the number density, $n[\rho] = \int \dd\oo \,\rho(\rr,\oo)$. By invoking eq~\ref{eq:mdft_bridge_decompo}, the FMT bridge functional can be written as

\begin{align}
      \Delta \Omega^\mathrm{FMT}_\mathrm{bridge}[\rho] &= \mathcal{F}^\mathrm{FMT}_\mathrm{exc}[\rho] - \mathcal{F}^\mathrm{FMT}_\mathrm{exc}[\rho_0] -\frac{\delta\mathcal{F}^\mathrm{FMT}_\mathrm{exc}}{\delta\rho(\rr,\oo)}[\rho=\rho_0] \Delta N[\rho] -  \Delta \Omega^\mathrm{FMT}_\mathrm{HRF}[\rho]
    \label{eq:mdft_bridge_decompo_fmt}
\end{align}
where

\begin{align}
    \Delta \Omega^\mathrm{FMT}_\mathrm{HRF}[\rho] &=   - \frac{k_BT}{2} \int\dd\rr\dd\rr' \Delta n(\rr) c^\mathrm{HS}(\rr, \rr') \Delta n(\rr')
\end{align}
with $c^\mathrm{HS}$ the direct correlation function of a hard-sphere fluid. In practice, we used the Carnahan-Starling approximation to construct the FMT functional because it gives a better equation of state than the Percus–Yevick approximation.~\cite{levesque_scalar_2012} The excess FMT functional (and the direct correlation function) depends on a single parameter, namely the hard-sphere radius $r_\mathrm{HS}$. This parameter was chosen so as to accurately reproduce the exact bridge functional for a homogeneous fluid (see the main text for details).

\subsection{Homogeneous fluid bridge}
\label{sec:si_hom_bridge}

For a homogeneous fluid (\textit{i.e.} independent of spatial and orientational degrees of freedom), the HRF functional is expressed as

\begin{align}
    \Delta \Omega^\mathrm{hom}_\mathrm{HRF} &= -\frac{k_BT}{2} \Delta n^2 \hat{c}(0) V
\end{align}
where 

\begin{align}
    \hat{c}(0) &= \frac{1}{N_\omega^2}\int\dd\oo\dd\rr'\dd\oo' c^{(2)}(\rr=0,\oo,\rr',\oo', n)
\end{align}
Conversely, the excess intrinsic free energy of a homogeneous fluid can be determined via isothermal integration,~\cite{hansen_theory_2006}

\begin{align}
   \mathcal{F}^\mathrm{hom}_\mathrm{exc}(n) = nVk_BT\int_0^n \dd n' \frac{1}{n'}\left( \frac{P(n')}{k_BTn'}-1 \right) 
   \label{eq:f_exc_p}
\end{align} 
Finally, the combination of eq~\ref{eq:mdft_bridge_decompo}, eq~\ref{eq:f_exc_p}, and the equation of state $P(n)$ yields the explicit expression of the bridge functional for a homogeneous fluid, expressed per unit volume:

\begin{align}
    \Delta\omega^\mathrm{hom}_\mathrm{bridge}(n) &\equiv \frac{\Delta\Omega^\mathrm{hom}_\mathrm{bridge}(n)}{V} = f^\mathrm{hom}_\mathrm{exc}(n) - f^\mathrm{hom}_\mathrm{exc}(n_0)   -\frac{\delta\mathcal{F}_\mathrm{exc}}{\delta\rho(\rr,\oo)}[\rho=\rho_0] \Delta n 
    + \frac{k_BT}{2} \Delta n^2 \hat{c}(0) \\
    f^\mathrm{hom}_\mathrm{exc}(n) &= nk_BT\int_0^n \dd n' \frac{1}{n'}\left( \frac{P(n')}{k_BTn'}-1 \right)
\end{align}
The bridge functional for a homogeneous fluid, expressed per unit volume, $ \Delta\omega^\mathrm{hom}_\mathrm{bridge}(n)$, is subsequently combined with the coarse-grained number density to construct the corresponding coarse-grained bridge functional (see the main text for details).

\section{Expansion onto generalized spherical harmonics}
\label{sec:si_generalized}

The one-body molecular density is a six-dimensional function that depends on both the translational (spatial) and rotational (orientational) degrees of freedom. 

\paragraph{Orientational degrees of freedom}

In practical implementations, the angular integrations are carried out on two distinct grids. In the first step, the molecular density is expanded in a basis of generalized spherical harmonics

\begin{align}
    \Delta \rho(\rr, \oo) = \sum_{m=0}^{n_\mathrm{max}}\sum_{\mu'=-m}^m\sum_{\mu=m}^m \sqrt{2m+1} \Delta\rho^m_ {\mu\mu'}(\rr)R^m_{\mu'\mu}(\oo)
\end{align}
where the generalized spherical harmonics are defined as

\begin{align}
    R^m_{\mu'\mu}(\oo) = r^m_{\mu'\mu}(\theta)\exp{\left(-\ii\mu'\phi-\ii\mu\psi\right)}
    \label{eq:generalized_spherical}
\end{align}
where \(r^m_{\mu'\mu}(\theta)\) denotes the generalized Legendre polynomial. Accordingly, the direct correlation function can be represented as an expansion in the basis of rotational invariants,

\begin{align}
    c(r, \oo_1, \oo_2, \hat{\rr}_{12}) = \sum_{m=0}^{n_\mathrm{max}} \sum_{n=0}^{n_\mathrm{max}} \sum_{l=\vert m-n\vert}^{m+n} \sum_{\mu=-m}^{m} \sum_{\nu=-n}^{n} c_{\mu\nu}^{mnl}(r) \Phi_{\mu\nu}^{mnl}( \oo_1, \oo_2, \hat{\rr}_{12})
\end{align}
where the rotational invariants are

\begin{align}
    \Phi_{\mu\nu}^{mnl}( \oo_1, \oo_2, \hat{\rr}) = \sqrt{(2m+1)(2n+1)} \sum_{\mu'=-m}^m \sum_{\nu'=-n}^n \sum_{\lambda'=-l}^l\begin{pmatrix} m & n & l \\ \mu^{'} & \nu^{'} & \lambda^{'} \end{pmatrix}  R^{m}_{\mu^{'}\mu}(\oo_{1}) R^{n}_{\nu^{'}\nu}(\oo_{2}) R^{l}_{\lambda^{'}0}(\hat{\rr}_{12})
\end{align}
where $\begin{pmatrix} m & n & l \\ \mu^{'} & \nu^{'} & \lambda^{'} \end{pmatrix}$ denotes the Wigner $3j$-symbol and $R^{m}_{\mu^{'}\mu}$ represents the generalized spherical harmonics (as defined in eq~\ref{eq:generalized_spherical}). \par

The orientational degrees of freedom are also decomposed into a grid of dimensions $N_\theta \times N_\phi \times N_\psi$, where the discrete angles $\theta_i, \phi_i, \psi_i$ are selected according to a Gauss quadrature scheme. The total number of angular grid points depends on $n_\mathrm{nmax}$ and is typically smaller than $2 n_\mathrm{nmax}$.
 
\paragraph{Spatial degrees of freedom}

\begin{figure}
\begin{center}
\includegraphics[width=0.7\columnwidth]{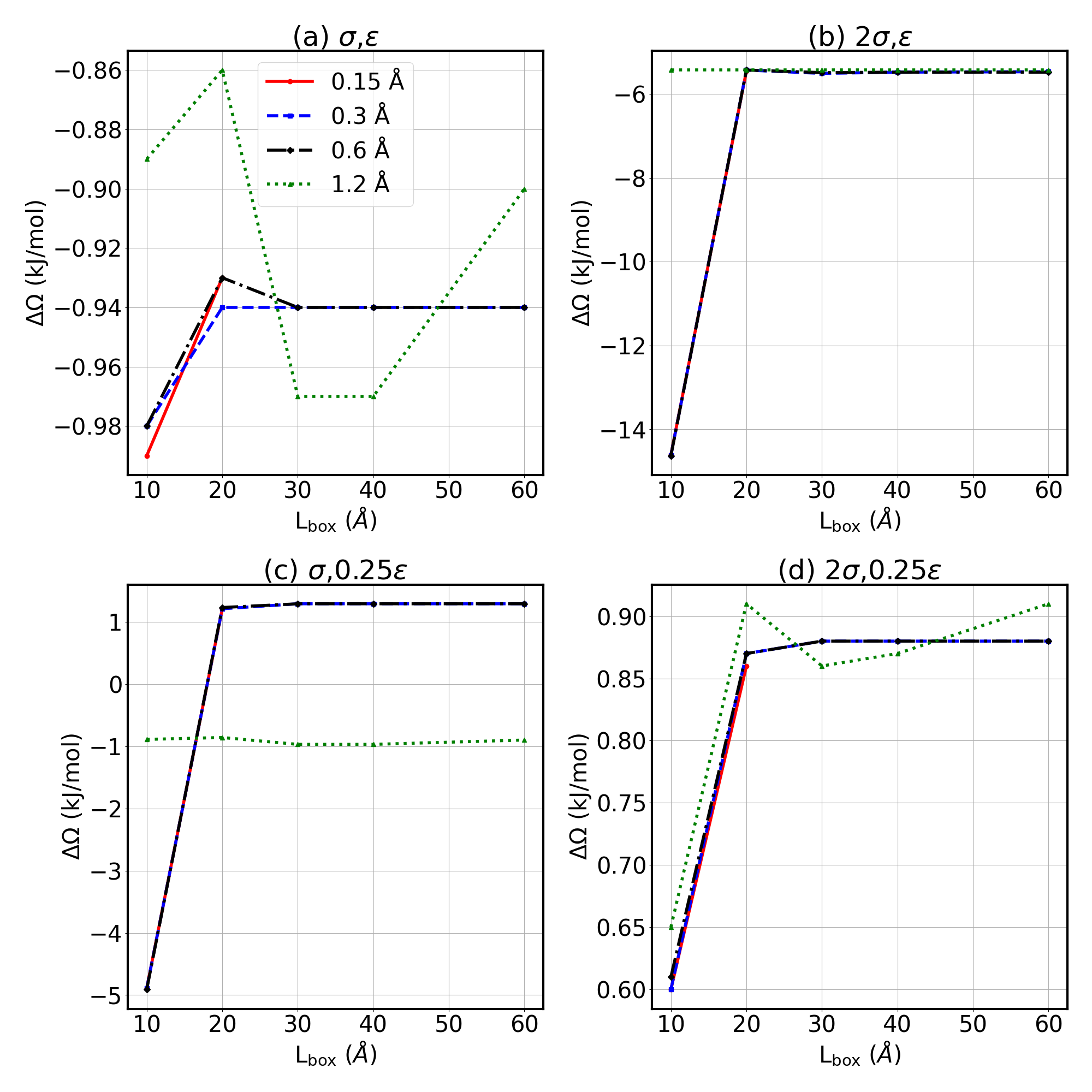}
\caption{Solvation free energies (SFE) computed with MDFT for different box sizes and different mesh sizes and for four different solutes of Lennard-Jones types, with sizes $\sigma_\mathrm{LJ}$ and interactions $\epsilon_\mathrm{LJ}$ indicated in each figure (with $\sigma = 3.405$~\AA~and $\epsilon = 1$~kJ/mol). }
\label{fig:sfe_vs_lbox}
\end{center}
\end{figure}

\begin{figure}
\begin{center}
\includegraphics[width=0.7\columnwidth]{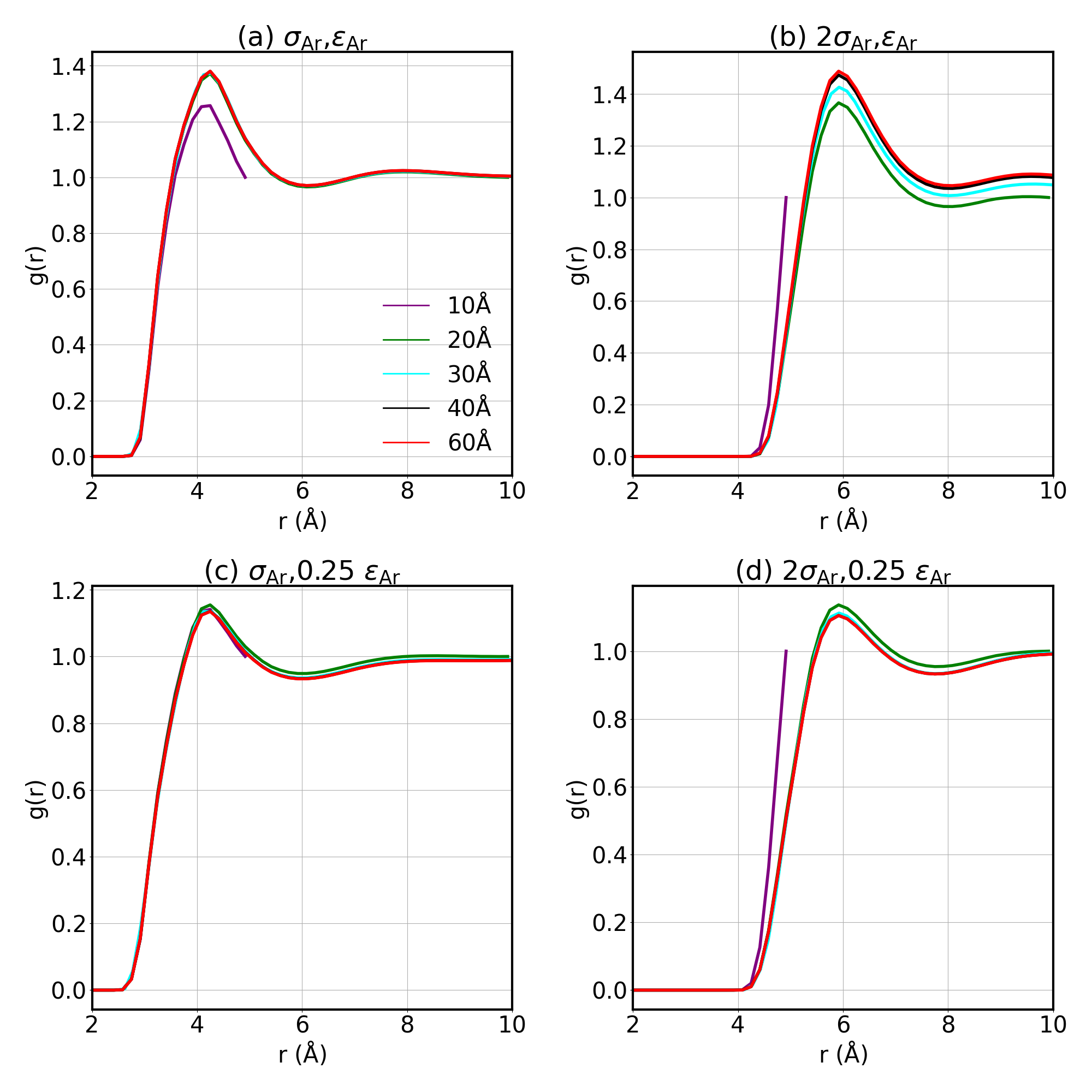}
\caption{Site-site radial distribution function LJ-C obtained with MDFT for different box sizes with a mesh size ($\Delta r = 0.3$~\AA) and for four different spherical solutes of Lennard-Jones types, with sizes $\sigma_\mathrm{LJ}$ and interactions $\epsilon_\mathrm{LJ}$ indicated in each figure (with $\sigma_\mathrm{Ar} = 3.405$~\AA~and $\epsilon_\mathrm{Ar} = 1$~kJ/mol).}
\label{fig:rdf_lj_vs_lbox}
\end{center}
\end{figure}

The spatial degrees of freedom are discretized on a three-dimensional grid comprising $N \times N \times N$ spatial nodes, where $N = L_\mathrm{box}/\Delta r$, $L_\mathrm{box}$ denotes the length of the cubic simulation box, and $\Delta r$ is the mesh size. We first assessed the convergence of thermodynamic and structural observables as function of both box size and mesh resolution. Figure~\ref{fig:sfe_vs_lbox} shows the solvation free energies (SFE) for four spherical solutes of different sizes and interactions ($\sigma_\mathrm{LJ} = \sigma_\mathrm{Ar}$ and  $2~\sigma_\mathrm{Ar}$, $\epsilon_\mathrm{LJ} = 0.25~\epsilon_\mathrm{Ar}$ and $\epsilon_\mathrm{Ar}$, with $\sigma_\mathrm{Ar} = 3.405$~\AA~and $\epsilon_\mathrm{Ar} = 1$~kJ/mol) characteristics as functions of the box length (ranging from 10~\AA\ to 60~\AA) and the mesh grid (0.15~\AA, 0.3~\AA, 0.6~\AA, and 1.2~\AA). Figure~\ref{fig:rdf_lj_vs_lbox} presents the carbon–solute and oxygen–solute radial distribution functions for the same set of spherical solutes as functions of the box length. Both the solvation free energies and the radial distribution functions were found to be converged for $L_\mathrm{box} = 30$~\AA\ and $\Delta r = 0.3$~\AA, corresponding to a grid of size $100 \times 100 \times 100$.

\newpage
\section{Radial distribution functions for the solutes}
\begin{figure}[h!]
\begin{center}
\includegraphics[width=0.5\columnwidth]{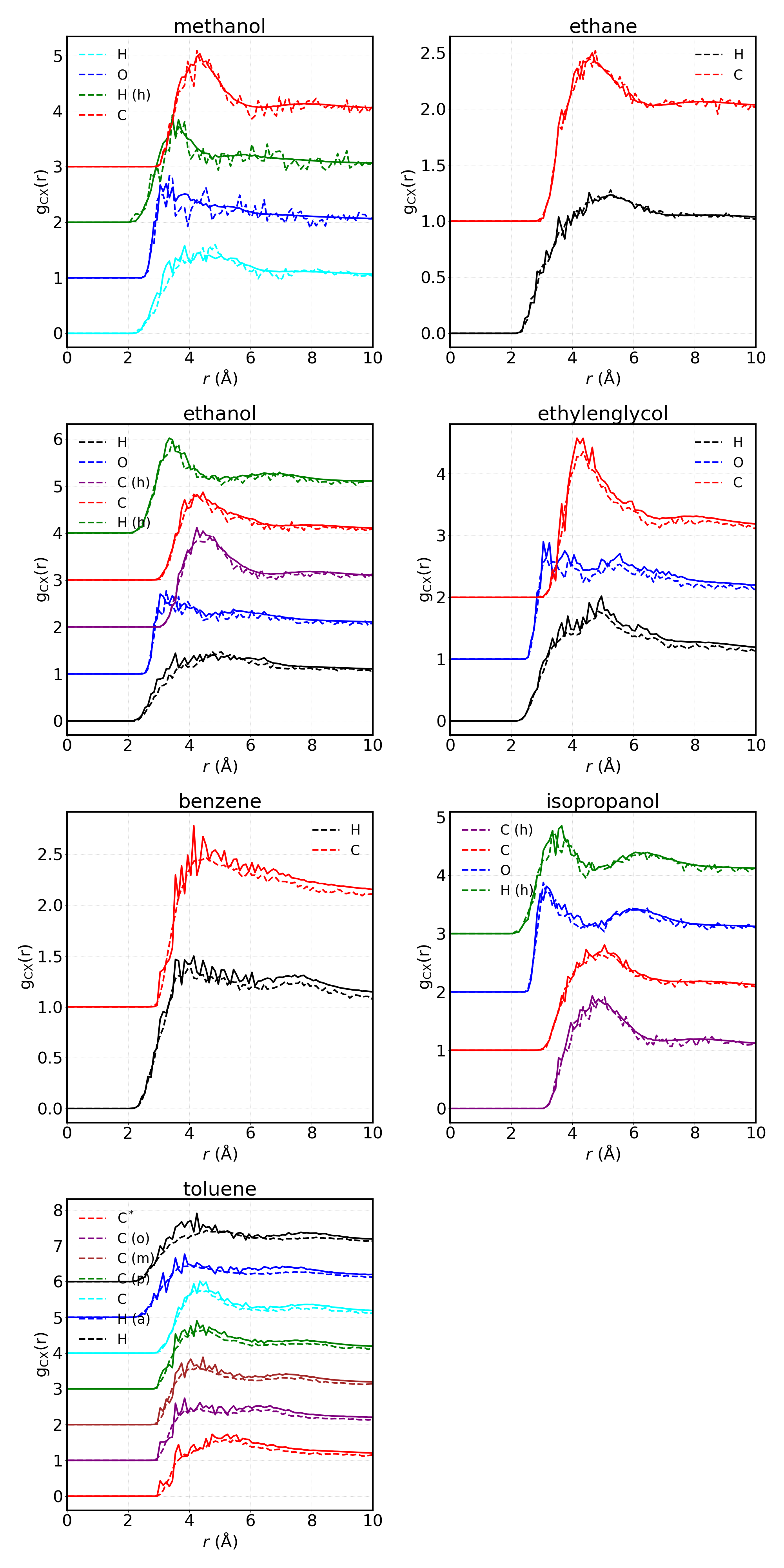}
\caption{Comparison of the solvation structures surrounding seven molecular solutes, represented through site–site radial distribution functions (RDF) between the carbon atom of CO$_2$ and all atoms of each solute. MDFT predictions are shown as solid lines, while MD simulation results are depicted as dashed lines. The RDF curves have been vertically shifted to facilitate interpretation of the figure. The label $(h)$ indicates hydrogen or carbon atoms bonded with an oxygen in a hydroxyl group. For toluene, the labels (*, o, m, p, a) correspond to the methylated, ortho, meta, para carbons and the aromatic hydrogens, respectively. }
\label{fig:rdf_mols_in_sco2}
\end{center}
\end{figure}

\newpage

\bibliography{biblio}

\end{document}